%% file: elsarticle-template.tex
\documentclass[final,5p,times,twocolumn]{elsarticle}

\usepackage{float}
\usepackage{graphicx}

\usepackage{amssymb}
\usepackage{url}
\usepackage{multirow}

\usepackage{scrextend}

\usepackage{kotex}

\usepackage{color}

\journal{arXiv}

\begin{document}

\begin{frontmatter}

\title{Watch Out! Smartwatches as criminal tool and digital forensic investigations}

\author{Seungjae Jeon}
\ead{jsj970515@dgu.ac.kr}

\author{Jaehyun Chung}
\ead{jaehyun226@gmail.com}

\author{Doowon Jeong\corref{cor1}}
\ead{doowon@dgu.ac.kr}

\address{College of Police and Criminal Justice, Dongguk University, Seoul, 04620, South Korea}
\cortext[cor1]{Corresponding author}

%

\begin{abstract}

In the rapidly advancing technological landscape, smartwatches have materialized as multifunctional devices integral to our daily routines. Smartwatches store a substantial amount of personal information, potentially serving as repositories of digital evidence. Thus, digital forensic researchers have devoted considerable effort to exploring smartwatch forensic techniques. However, it has been observed that prior studies have primarily treated smartwatches as mere storage mediums for digital evidence, neglecting their potential role in criminal activities. This paper presents the information leakage perpetrated through smartwatches. We represent crime scenarios in an environment where smartphones are not available, considering that the perception that smartphones can be used as tools for criminal behavior prevails in many organizations, while the potential of similar-use smartwatches is often overlooked. We detail mechanisms for information leakage via file transfer and camera control using smartwatches. Additionally, we present methods to investigate each crime incident through smartwatch forensics. Finally, we describe the limitations of post-incident responses and propose proactive measures to prepare for potential crimes involving smartwatches. 

\end{abstract}


\end{frontmatter}


\input{_sections/sec_1-Introduction.tex}
\input{_sections/sec_2-Related_Works.tex}
\input{_sections/sec_3-Background.tex}
\input{_sections/sec_4-Smart_Watch_As_Criminal_Tool.tex}
\input{_sections/sec_5-Forensic_Methods.tex}
\input{_sections/sec_6-Discussion.tex}
\input{_sections/sec_7-Conclusion.tex}





\begingroup
\raggedright

\bibliography{sample.bib}
\bibliographystyle{unsrt}
\endgroup

\end{document}

%% file: _sections/sec_1-Introduction.tex
\section{Introduction}
\label{sec:introduction}
\begin{table*}[h]
    \begin{center}
    \caption{Smartwatch information used in the experiment}
    \label{tab:Smartwatch information used in the experiment}
    \begin{tabular}{llllll}\hline
Product Name       & Model Number & \begin{tabular}[c]{@{}l@{}}System Version\\ (Android Version)\end{tabular} & Wear OS Version & CPU abi & ADB Host Name     \\ \hline
Galaxy Watch 5 Pro & SM-R920      & 11                                                                        & 3.5             & armeabi-v7a  & projectxbl\\
Galaxy Watch 5     & SM-R910      & 11                                                                         & 3.5             & armeabi-v7a & heartbl\\
Galaxy Watch 4     & SM-R860      & 11                                                                         & 3.5             & armeabi-v7a & freshbs\\ \hline
\end{tabular}
\end{center}
\end{table*}

The unauthorized leakage of industrial technology and business information constitutes a grave crime, acting as a fatal threat to the survival of majority corporations. According to a report by the National Intelligence Center for Industrial Security, there were a total of 117 detected instances of overseas leakage of industrial technology from 2017 to 2022. Among these, the leakage of national core technologies accounted for 36 cases (approximately 30.7\%), with the estimated damage amounting to 26 trillion KRW~\cite{article3}. In response to this phenomenon, the Korean National Police Agency initiated a 'Special Crackdown on Crimes Threatening Economic Security' in February 2023. The interim results reveal that 35 cases of information leakage crimes were investigated over a four-month period, with 30 cases (approximately 85.7\%) being attributed to insiders, and 5 cases (approximately 14.3\%) to outsiders, confirming that the primary source of information leakage is within corporations~\cite{police_press_release}. 
In respond, there has been a research conducted through constructing information leakage scenarios and analysing the key indicators derived from those~\cite{park2020study}. The present study, inspired by the methodology of previous research, proposes a scenario that utilizes smartwatches, one of the most popular and easily portable IoT(Internet of Things) devices, as a new medium for storing and transmitting internal information leakage.

In Korea, there has been a systematic rise in the adoption of smartwatches across diverse age group, resulting in an increase from 12.0\% in 2020 to 19.0\% in 2021, reaching 24.0\% in 2022~\cite{korea_gallup_stats}. These statistics confirm smartwatches standing out as a prevalent IoT (Internet of Things) tool within the domestic technological landscape. Traditional digital forensic research has predominantly centered on situations wherein the suspect or victim incidentally wore a smartwatch, leveraging the embedded data as evidential support for specific allegations. This inclination stems from the multifaceted capabilities of smartwatches, encompassing the generation of exercise-related data, which allows for the discernment of an individual's health parameters and GPS locations. Moreover, the synchronization feature with smartphones enables users to view appointments and notifications directly on their smartwatch, enhancing its applicability in legal and criminal contexts. 

This study differ from prior research by concentrating on smartwatches as potential evidence for leaking confidential information. Substantial normative and technical efforts have been made to regulate the use of high-functioning communication tools like smartphones due to concerns over criminal misuse, however corresponding to it, regulation against smartwatches is conspicuously lacking. Focusing on this observed disparity in security policies, this study endeavors to elucidate the potential risks of criminal activities involving smartwatches and seeks to propose specific forensic countermeasures in response to these threats.

The remainder of the paper is organized as follows: we delve into the existing body of knowledge related to smartwatch forensics and present cases where smartwatch analysis has been instrumental in apprehending criminals in Section~\ref{sec:related works}. 
Section~\ref{sec:Background} examines the characteristics and specifications of Wear OS. 
Section~\ref{sec:criminal tool} explores the feasibility of committing crimes using smartwatches. 
Section~\ref{sec:forensic methodos} introduces a forensic analysis approach to examine data generated by smartwatches. 
Then, in Section~\ref{sec:Discussion}, we discuss the limitations of the forensic methods outlined in Section~\ref{sec:forensic methodos} and propose strategies for preventing smartwatch-related crimes.
Finally, Section~\ref{sec:conclusion} concludes this paper.

%% file: _sections/sec_2-Related_Works.tex
\section{Related works}
\label{sec:related works}
Several studies have investigated the forensic analysis of smartwatches, with a focus on different operating systems and models. Ruthani and Dahiya~\cite{rughani2015analysis} conducted research on artifacts from the Android Wear operating system smartwatch. The Android Wear operating system serves as the predecessor to the Wear OS, now equipped in the latest smartwatches, and connects with devices running Android 4.3 or higher. When the device is linked with a smartphone, the watch itself updates various data from the smartphone, including Google Mail, calendar entries, mobile phone notifications, and etc. This study utilized the "dd" command to dump the smartwatch's image, facilitating an analysis of the updated information within the device using artifacts.  This analytical process revealed connected device specifications, the history of voice commands, logs of notifications, and DropBox artifacts.

Odom et al.~\cite{odom2019forensic} analyzed Samsung's Galaxy Gear S3 using the Tizen OS and Apple's Apple Watch Series 3 using the Watch OS. Contact, schedule, alarm, reminder, password, email, multimedia, phone history, SMS, IM, and voice commands remaining inside the smartwatch were compared with the bluetooth connection method and the Stand-Alone method. As a return of the analysis, the study produced a Galaxy Gear S3 analysis tool called Gear Gadget.

Kim et al.~\cite{kim2021security} conducted an analysis of Samsung's Galaxy Gear S3, Apple's Apple Watch 5, and Garmin Vivosport, and proposed a forensic model. The study proposed a forensic model and employed various forensic methods, such as logical extraction through PC connections and hardware-based methods like PCB service port, PCB debugging port, and chip-Off. In particular, the focus was on analyzing device information and health care information.

Previous studies have contributed to the field of smartwatch forensics by analyzing the data stored within these devices and highlighting their potential as sources of evidence. Notably, real-life cases have emerged where smartwatch data played a crucial role in criminal investigations in particular homicide crimes. In the case of the 2021 bridal murder in Greece, the husband claimed his bride was killed because her honeymoon house was robbed. However, as a result of the bride's smartwatch forensics, the bride's heartbeat was also recorded after the time of her husband's testimony, which was used as evidence of her husband's perjury and proof of charges~\cite{article2}. Similarly, in 2022, a sudden increase in heart rate and stride recorded on the smartwatch of victim Police Community Support Officer Julia James was presented as evidence to prove when she found the culprit~\cite{article1}.

Previous studies demonstrate that smartwatches are devices that have very close information with users, such as smartphones, and actual cases prove the contribution of previous studies.  However, some notable changes have occurred. Google's Android Wear, once the primary focus of analysis, has been succeeded by Wear OS. Concurrently, Tizen OS, the operating system of Galaxy Gear S3, has seen a decline in market share since Samsung adopted Wear OS Powered by SAMSUNG for the Galaxy Watch 4 and 5. Subsequently, in the case of Apple's Apple Watch, as the deletion of the PC connection port from the 7th generation, logical analysis through the connection between the smartwatch and the PC was not possible.

In order to advance existing research in consideration of this situation, this study concentrates on the latest wearable OS, Wear OS Powered by SAMSUNG. Moreover, while previous studies primarily assumed that smartwatches were incidental to specific crimes, this research explores scenarios where smartwatches are directly employed as tools for criminal activities, emphasizing the potential risks associated with smartwatch misuse.

%% file: _sections/sec_3-Background.tex

\section{Background}
\label{sec:Background}

Wear OS Powered by SAMSUNG is a smartwatch operating system developed in collaboration between Samsung and Google. Specifically, it is the designated OS for the Galaxy Watch 4 and Galaxy Watch 4 Classic, both of which were released in 2021, as well as for the Galaxy Watch 5 and Galaxy Watch 5 Pro, introduced to the market in 2022. Wear OS Powered by SAMSUNG incorporates the One UI Watch interface, which provides a familiar user experience for users transitioning from the previous Galaxy Watch 3. Notably, being built on Wear OS enables the utilization of the Android Debug Bridge (ADB) for various tasks and analysis.
\begin{figure*}
\begin{center}
\includegraphics[width=0.95\linewidth]{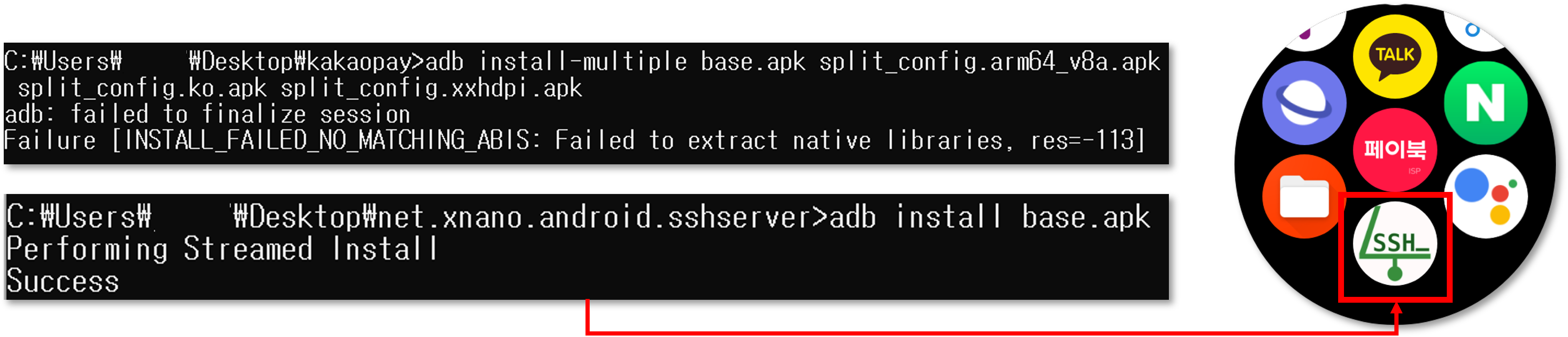}
\end{center}
\caption{Smartphone application SSH Server installed on the smart watch}
\label{fig:adb_install}
\end{figure*}

Table~\ref{tab:Smartwatch information used in the experiment} presents an overview of the software information and ABI (Application Binary Interface) versions of settings for smartwatches using the Wear OS Powered by SAMSUNG operating system. On Android smartphones, the part marked as `Android version' was marked as `System version' on the smartwach, so it was necessary to check whether the system is indeed Android. This was verified by the ADB command \texttt{`getprop ro.build.version.release'} confirming that the system is Android, and it was verified that smartwatches using Wear OS Powered by SAMSUNG operate on Android 11. In addition, the ADB command \texttt{`getprop ro.product.cpu.abi'} confirmed that each smartwatch uses armeabi-v7a. Abi, or the Android Application Binary Interface, shows that the Galaxy Watch 4 and 5 series operate on a 32-bit architecture based on ARMv7.

%% file: _sections/sec_4-Smart_Watch_As_Criminal_Tool.tex
\section{Smart watch as criminal tool}
\label{sec:criminal tool}

In the realm of smartphones, corporations foresaw the possibility of malicious exploitation and instituted various levels of security measures, ranging from superficial procedures like the affixation of security stickers to more comprehensive preventive actions such as prohibiting the devices' entry. This has been further fortified with technological strategies, such as Mobile Device Management (MDM), to safeguard organizational integrity. However, compared to smartphones, the potential abuse of smartwatches has been relatively underestimated. Consequently, the security regulations related to smartwatches are not as strict as those for smartphones. Reflecting this situation, this section explains that while security regulations for smartphones prevent their use as tools for information leakage, the lack of security regulations for smartwatches allows for the possibility of insider crimes where the role of smartphones can be replaced with smartwatches.

\begin{figure}[h]
\begin{center}
\includegraphics[width=0.48\textwidth]{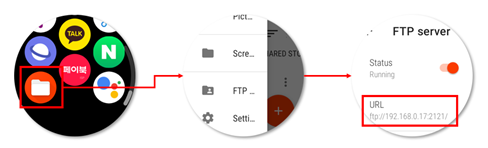}
\end{center}
\caption{A file explorer application for smartwatches that supports FTP}
\label{fig:file_explorer}
\end{figure}

\begin{figure*}[h]
    \begin{center}
    \includegraphics[width=0.98\linewidth]{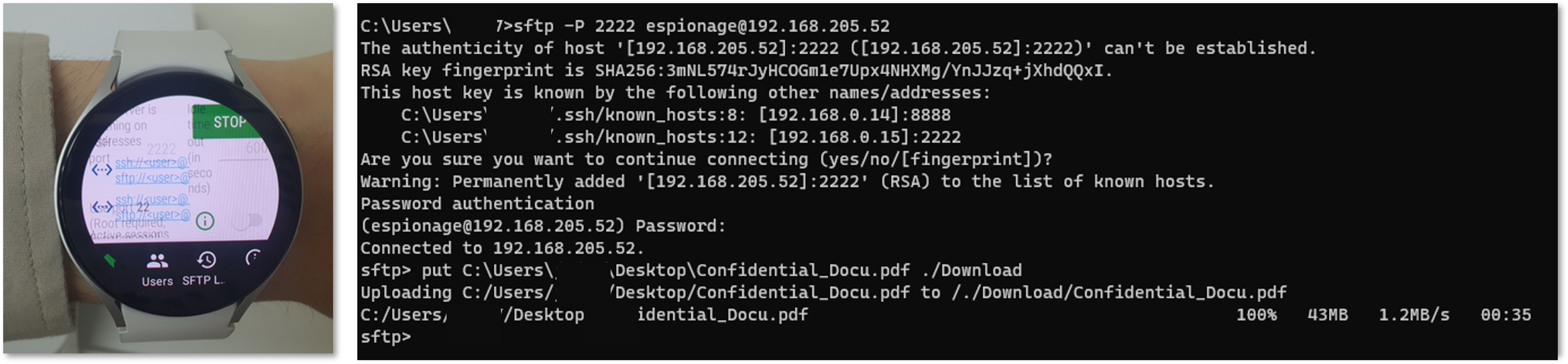}
    \end{center}
    \caption{SSH connection and file transfer between smartwatch and PC}
    \label{fig:using_ssh}
\end{figure*}

\subsection{Using official smart watch application}
\label{subsec:official smartwatch applications}

In general, Wear OS Powered by SAMSUNG smartwatches can install applications made for smartwatches by browsing and selecting from Google Play Electronic Service Distribution (ESD). While Most applications are made for general use of smartwatches, such as scheduling and health care, some smartwatches internal storage explorer applications provide FTP functionality. An example of such an application is File Explorer FTP Server (\texttt{com.corproxy.files}), which can be found on the Google Play Store. This commercial application enables file exploration capabilities on smartwatches and provides an FTP server function. By engaging the FTP server functionality on a smartwatch, users are able to establish a wireless connectivity channel between the smartwatch and a personal computer through a Wi-Fi network. This is achieved without the necessity for an external port for direct PC linkage, expanding the utility and flexibility of the smartwatch's interface capabilities.

Figure~\ref{fig:file_explorer} illustrates the utilization of File Explorer FTP Server to initiate a FTP server in smartwatch. Users can connect to the FTP server on their smart watch using PC applications such as FileZilla and FTP 7, and user can also connect their PC and smartwatch by entering the FTP server address and port number in the file explorer path window. The presence of File Explorer FTP Server on the publicly accessible Google Play Store heightens the concern regarding its potential misuse for unauthorized data extraction. Nonetheless, even individuals without expertise in Android-related technologies can effortlessly transfer information from a PC to a smartwatch.

\subsection{Using forced installation of smartphone application}
\label{subsec:forced installaion smartphone applications}

The utilization of File Explorer FTP Server for unauthorized file extraction offers broad accessibility, as it leverages commercial applications. However, a crucial limitation arises from its reliance on FTP, a protocol that does not facilitate encrypted communication, thereby presenting an inherent risk to potential attackers. Additionally, since the application market for smartwatches is mainly focused on health care and calendar management, it is challenging to locate applications equipped with features that would be conducive to malicious intentions. Moreover, the fact that the installation history remains within the user's Google Play account adds the complexity of evidence elimination thereby posing a challenge to any attempts at concealing unauthorized activities.

These constraints can be circumvented by directly deploying smartphone applications onto the device, thus eliminating the dependence on the existing smartwatch application ecosystem. As previously delineated, the smartwatch confirmed that the OS is based on Android 11 and ADB can be used. This means that it is possible to install an APK file for a smartphone or a self-produced smartwatch application on a smartwatch through the \texttt{`adb install'} command, without necessitating reliance on Google Play to install an application for a smartwatch.

Figure~\ref{fig:adb_install} demonstrates the results of utilizing using the \texttt{`adb install'} command to install APK files of a 64-bit application, KakaoPay (com.kakaopay.app), and a 32-bit application, SSH Server (net.xnano.android.sshserver), specifically targeting the Galaxy Watch 5. Due to the fact that Galaxy Watch 4 and 5 series both operate on an ARMv7-based 32-bit architecture, installation of 64-bit APK file resulted in an ABI matching error. On the other hand, the 32-bit APK file was verified to be installed on the smartwatch without any issues.

\begin{figure}[h]
\begin{center}
\includegraphics[width=0.48\textwidth]{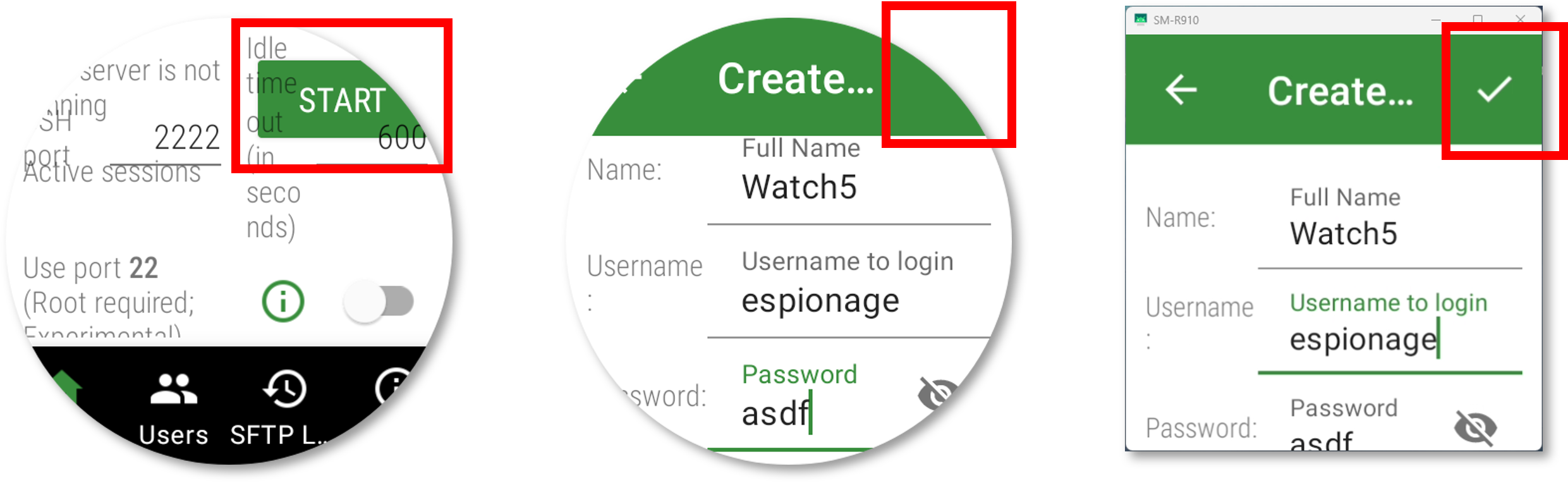}
\end{center}
\caption{Problems on the screen when installing smartwatches for smartphone applications}
\label{fig:screen}
\end{figure}

However, installing smartphone applications on smartwatches can lead to graphical issues. Applications for smartphones are made assuming a screen that is larger and rectangular than a smartwatch. Therefore, as shown in the picture on the left side of Figure~\ref{fig:screen}, applications for smartphones installed on smartwatches may have screen components that are overllaped. In addition, applications for smartphones are assumed to be designed for a rectangular display, while Samsung smartwatches have a circular screen. Therefore, as compared with the middle image and the right image of Figure~\ref{fig:screen}, there may be instances where the elements at the vertices of the rectangular display are not visible within the boundaries of the smartwatch's circular screen. Nevertheless, these visual issues do not necessarily render the application unusable.

In Figure~\ref{fig:adb_install}, base.apk is an apk file of SSH Server (\texttt{net.xnano.android.sshserver}), and the application for the smartphone functions to use the SFTP protocol by opening the ssh server on the Android smartphone.
Figure~\ref{fig:using_ssh} demonstrates the operation of the SSH Server application on the smartwatch and the SSH connection between the smartwatch and the PC. Although there are issues with overlapping screen elements, these can be overcome by manipulating the application's settings. A method to establish a connection between a personal computer (PC) and a smartwatch involves utilizing the Secure File Transfer Protocol (SFTP) through the Windows command prompt. By executing the command \texttt{`sftp -P [SSH server port number] [user name]@[smartwatch IP]'} within the command prompt as shown on the right side of Figure~\ref{fig:using_ssh}, a secure channel can be established between them. This illustrate how an individual can easily transfer of data from the PC to the smartwatch using the \texttt{`put'} command within the SSH shell, potentially leading to unathorized data leakeage.

\subsection{Smartwatch as a hidden camera control device}
\label{subsec:using hidden cameara}

\begin{figure*}[h]
    \begin{center}
    \includegraphics[width=0.95\textwidth]{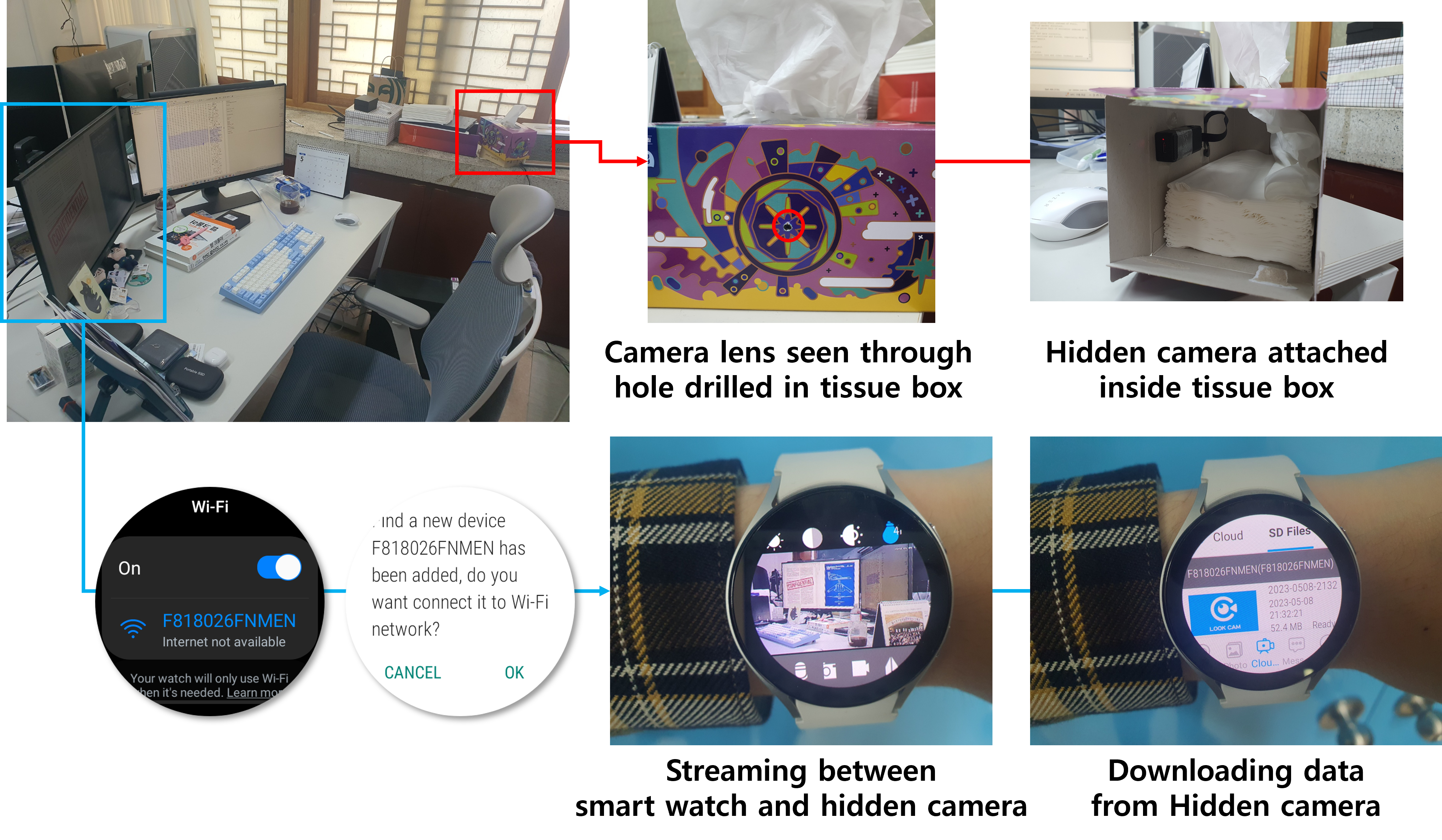}
    \end{center}
    \caption{Environment when connecting a hidden camera to a smart watch}
    \label{fig:hidden_camera_env}
\end{figure*}

Galaxy Watch 4, 5 series does not feature an integrated camera. However, by installing a compact camera control application designed for small cameras on the smartwatch, there is a possibility of utilizing a smartwatch as a control device for hidden cameras. Notably, the Lookcam application (package name: \texttt{com.view.ppcs}) serves as an illustrative example supporting this hypothesis.

Recently, small cameras have been releasing Wi-Fi to enable wireless connections with controls. Figure~\ref{fig:hidden_camera_env}   illustrates a scenario in which a small camera, clandestinely smuggled into an office and disguised as a tissue box, is connected to a smartwatch to leak information. Some companies may only conduct security inspections on items that employees are taking out of the office at the end of the day, neglecting to scrutinize items brought into the office at the start of the day. 

By receiving Wi-Fi emited by hidden cameras concealed within an innocuous object such as a tissue box, from outside the office, it was possible to remotely control hidden cameras in the office through a smartwatch. Since not only video recording but also voice recording functions are supported, the utilization of hidden cameras in close proximity to individuals with access to sensitive technology and management information, or within conference rooms, poses a significant risk as the captured data can be surreptitiously transmitted via the smartwatch. In such a controlled environment, smartwatches can act as a security threat because they have sufficient performance to replace the functions of smartphones. By exploiting these security procedures, a leaker can bring in a small camera into an environment and then connect it to a smartwatch via the camera's Wi-Fi signal, allowing for remote control over the camera. This fact makes it possible for the leaker to periodically link the camera to stream or download videos event without a smartphone. If the data is accumulated enough, the leaker could dismantle or dispose the small camera that was hidden in the company, avoiding detection during the security checks on outgoing items. The information could be safely stored in the smartwatch's memory and sneakingly exported outside the premises.

\begin{figure}[h]
\begin{center}
\includegraphics[width=0.48\textwidth]{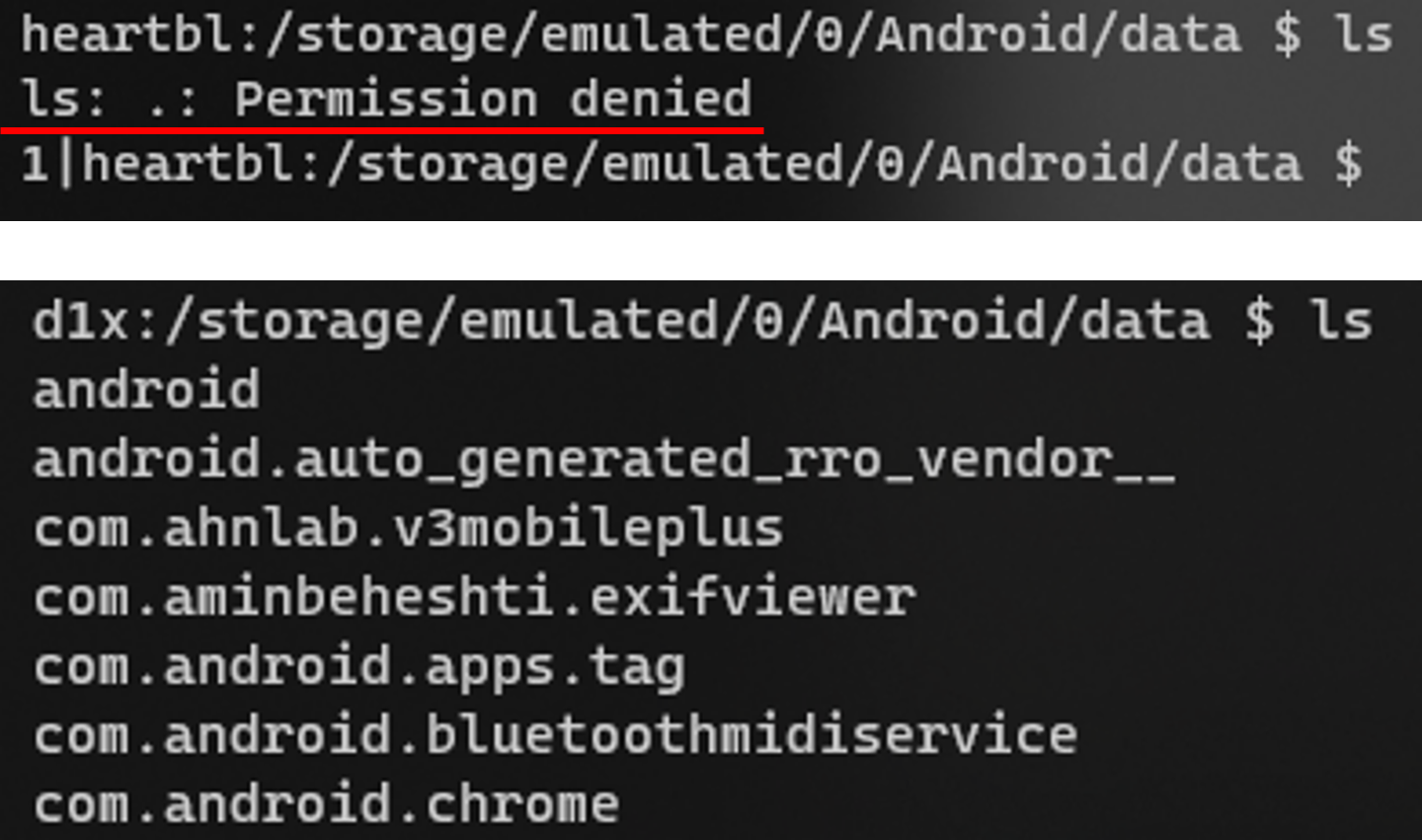}
\end{center}
\caption{Galaxy Watch 5 (top) / Galaxy Note 10 (bottom)}
\label{fig:compare_phone_and_watch}
\end{figure}

It is important to note that these findings highlight the security implications associated with smartwatches and emphasize the need for robust security measures to safeguard against unauthorized access and potential privacy breaches. Furthermore, the installation and use of applications should be regulated and monitored to prevent misuse and mitigate the risks posed by hidden camera control via smartwatches.

%% file: _sections/sec_5-Forensic_Methods.tex
\section{Forensic Methods}
\label{sec:forensic methodos}

\subsection{The difference between smart watch forensics and smartphone forensics}

Galaxy Watch 4 and 5, utilizing Wear OS Powered by SAMSUNG, present two main differences that complicate the application of traditional mobile forensic research and investigative methods. The first difference is the absence of a USB port. Since modern smartwatches primarily use wireless charging, many of them do not include a USB port. Consequently, traditional smartphone forensics, which rely on using a wired connection to a PC through the USB port  and then employing mobile data acquisition tools such as MD-NEXT for imaging and analyzing the device's data, cannot be applied. Additionally, the lack of a USB port creates differences in the accessible paths. As of Android 11, the path /storage/emulated/0/Android/data is inaccessible via the device's default  search function and can only be accessed via a wired connection to a PC. Thus, as shown in Figure ~\ref{fig:compare_phone_and_watch}, in the case of Wear OS Powered by SAMSUNG, which is based on Android 11 and only allows wireless connections, access to this path is impossible, which sets them apart from Android smartphones.

The second difference is the absence of publicly available firmware. The application package data in the /data/data path is a primary target for analysis in Android smartphone forensics. Accessing this path requires Super User (SU) permissions. In Android smartphone investigations, Full File System imaging can be used to view the data in this path without SU permissions using  forensic programs. However, this method is inapplicable for smartwatches, as previously described. 
Moreover,  during research, experiments may be conducted on a pre-rooted Android device, and SU permissions can be obtained in ADB Shell to access the data in the /data/data path. With the Galaxy Watch 4 and 5, which use the Wear OS Powered by SAMSUNG, rooting is currently challenging as the firmware is not publicly available. This leads to another significant difference from Android smartphones in that access to the /data/data path is unavailable, restricting the ability to analyze the application package data.

Reflecting these differences, an appropriate approach for analyzing smartwatches is the Android live forensic method using ADB. As was the case with installing smartphone APK files on a smartwatch, using ADB's wireless debugging feature enables a connection between an investigator's PC and the smartwatch under examination. Although obtaining Super User  permissions is not feasible, it is possible to transfer existing files from an Android smartwatch to a PC within the allowed permissions. Most importantly, this method enables user behavior analysis utilizing various logs provided by ADB, such as dumpsys logs, specifically designed for developers.

\subsection{Digital Evidence From Smart Watch}
\label{subsec:digital_evidence_from_smart_watch}

\begin{figure*}[hp]
    \begin{center}
    \includegraphics[width=0.95\linewidth]{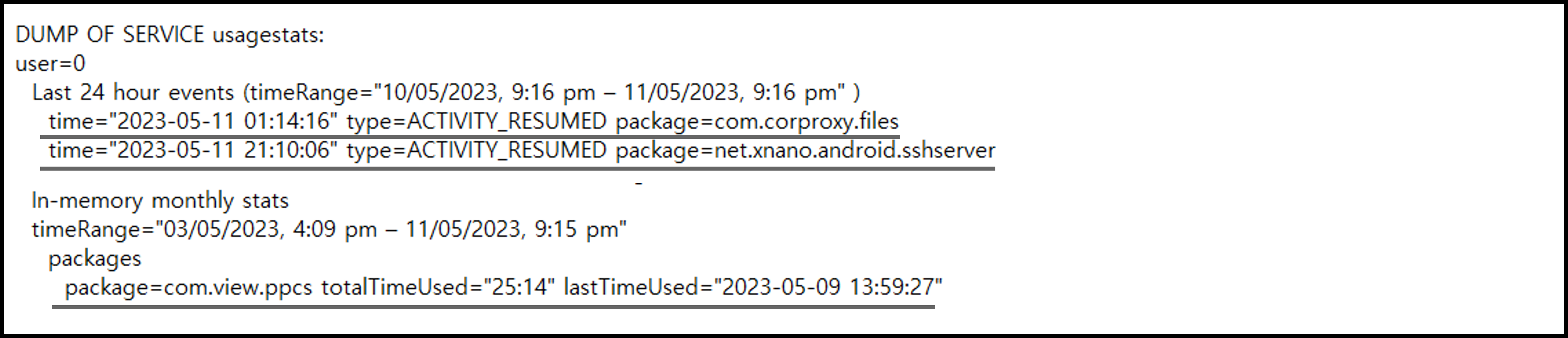}
    \end{center}
    \caption{dumpsys usagestats}
    \label{fig:usagestats}
\end{figure*}

\begin{figure*}[hp]
    \begin{center}
    \includegraphics[width=0.95\linewidth]{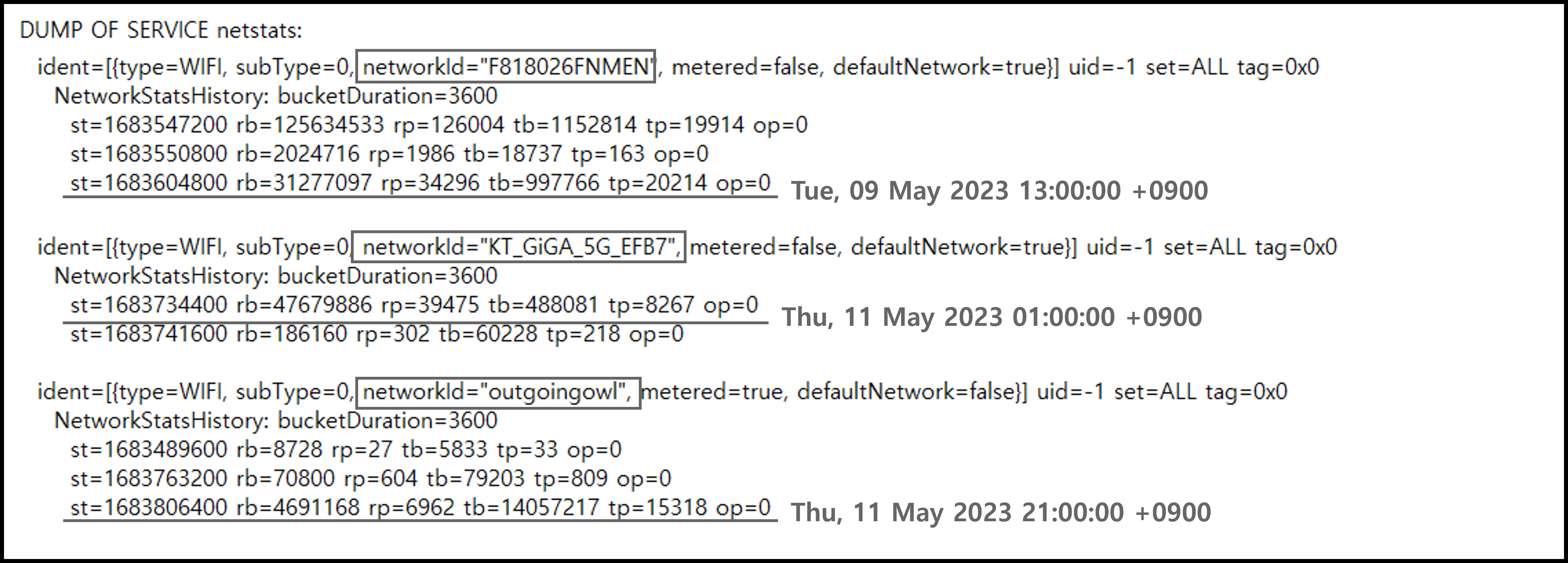}
    \end{center}
    \caption{dumpsys netstats}
    \label{fig:netstats}
\end{figure*}

\begin{figure*}[hp]
    \begin{center}
    \includegraphics[width=0.95\linewidth]{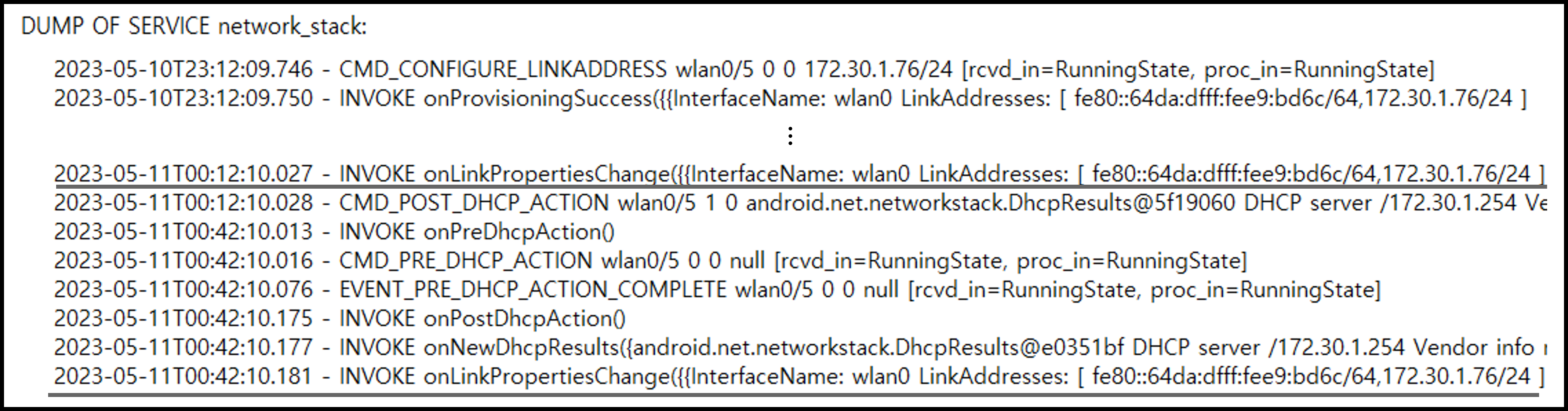}
    \end{center}
    \caption{dumpsys network\_stack(2023-05-11 09:56)}
    \label{fig:network_stack1}
\end{figure*}

\begin{figure*}[hp]
    \begin{center}
    \includegraphics[width=0.95\linewidth]{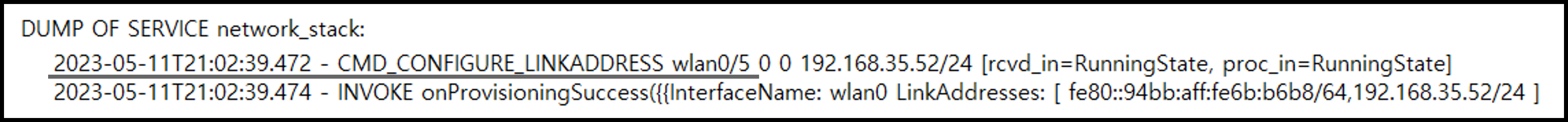}
    \end{center}
    \caption{dumpsys network\_stack(2023-05-11 21:45 )}
    \label{fig:network_stack2}
\end{figure*}

\subsubsection{dumpsys usagestats}
\label{subsubsec:dumpsys_usagestats}

Several previous studies have highlighted that dumpsys, an Android sdk, can be useful as a forensic method
~\cite{bortnik2021android, cheng2018sedalvik, easttom2021methodology}. 
This also applies  to Galaxy Watch 4 and 5. The Android 11 device stores usagestats in the path \texttt{/data/system\_ce/usagestats/0}, allowing investigators to examine the user's app usage history when analyzing the artifact~\cite{Kang2021android, kwon2021user}. Previous research analyzed this usagestats information by extracting files located on the path from Android smartphones.

Accessing the \texttt{/data/system\_ce/usagestats/0} path requires superuser privileges, which necessitates rooting the device for extraction and analysis. But relevant information can be examined without superuser privileges by using the dumpsys usagestats command in the ADB Shell. Figure~\ref{fig:usagestats} presents a portion of the usagestats data obtained using the dumpsys usagestats command. Up to 24 hours before the command is executed, logged  information corresponding to the type, such as  \texttt{FOREGROUND\_SERVICE\_START}, \texttt{ACTIVITY\_RESUMED}, and \texttt{NOTIFICATION\_INTERRUPTION}, provides insight  which applications were brought to the foreground of the smartwatch, and which application sent  notification~\cite{android-studio-docs}. App operation information for one week, one month, and one year can also be checked, but unlike 24h usagestats, precise information in hours, minutes, and seconds cannot be read , although  information such as the time frame the device was last used, and number of times the device was used  can be examined.

\subsubsection{dumpsys netstats}
\label{subsubsec:dumpsys_netstats}
The dumpsys netstats command provides valuable insights for investigators regarding the Wi-Fi network information accessed by the smartwatch. Figure~\ref{fig:netstats} is the result of using the command. First, the ‘networkId' item represents the name of the connected Wi-Fi network, and ‘st' represents the connected time. It is important to note that, ‘st’ is expressed in units of one hour. Therefore, for example, st appears as 1683718800 (20:00 on 10 May 2023) even if it was connected for only 10 minutes from 20:40 to 50 on 10 May 2023. ‘rb' and ‘rp' represent the size and number of packets received, and ‘tb' and ‘tp' represent the size and number of packets transmitted.

\subsubsection{dumpsys network\_stack}
\label{subsubsec:dumpsys network_stack}

Figure~\ref{fig:network_stack1} and Figure~\ref{fig:network_stack2} present the result of the dumpsys network\_stack command. This command offers information about network interfaces, network topologies, and network stacks, and enabling investigators to examine  the data exchanged with the Wi-Fi network through the DHCP protocol. Therefore, the dumpys network\_stack command allows the network interface of the smartwatch recorded over time to determine the private IP address given by the Wi-Fi network. However, it is important to note that when the smartwatch is rebooted, the existing record is replaced by a new record devoid of any former data , making it impossible to access the previous network stack information.

\begin{figure*}
\begin{center}
\includegraphics[width=0.95\textwidth]{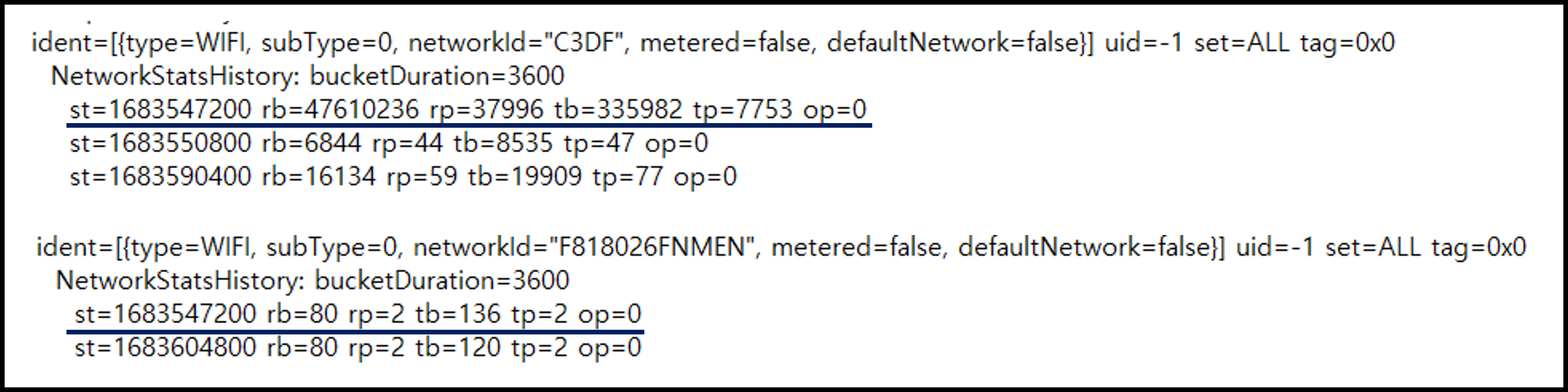}
\end{center}
\caption{Two networkId with the same st value}
\label{fig:same_st}
\end{figure*}

\subsection{Case Study}
\label{subsec:Case_Study}

By combining the information obtained from usagestats, netstats, and network\_stack through ADB, it becomes apparent that the user established a connection between their smartwatch and PC using the FTP protocol, followed by a subsequent connection using the SFTP protocol. The smartwatch was then utilized as a control device for hidden cameras.

\subsubsection{Connection between Smart watch and PC using FTP Protocol}
\label{subsubsec:find_ftp}

First, looking at the usagestats log , as seen in Figure~\ref{fig:usagestats}, the user started the File Explorer FTP Server (\texttt{com.corproxy.files}) application at 01:14:16 on May 11, 2023. Subsequently, the netstats records in Figure~\ref{fig:netstats} indicate that \texttt{KT\_GiGA\_5G\_EFB7} was the name of the Wi-Fi used by the smartwatch for FTP connection during that time, and about 47MB worth of data  was downloaded. 
As seen in Figure~\ref{fig:network_stack1}, network\_stack log shows that the private IP given when connecting with the corresponding \texttt{KT\_GiGA\_5G\_EFB7} was 172.30.1.76. 
By cross-verifying the information obtained from the smartwatch and the evidence left on the PC, the connection information with the PC can be confirmed. For example, if a user uses FileZilla, a typical FTP and SFTP protocol connection application, the contents of \texttt{filezilla.xml} and \texttt{recentservers.xml} files located in the \texttt{\%UserProfile\%\textbackslash AppData\textbackslash Roaming\textbackslash FileZilla} path can be compared to the system log record of the smartwatch.

\subsubsection{Connection between Smart watch and PC using SFTP Protocol}
\label{subsubsec:find_sftp}

Referring to the usagestats, as seen in Figure~\ref{fig:usagestats}, it can be observed that the user launched the SSH Server (\texttt{net.xnano.android.sshserver}) application at 21:10:06 on May 11, 2023. Subsequently, netstats records, as seen in Figure~\ref{fig:netstats}, indicate that the Wi-Fi used by smartwatches for SFTP connections was \texttt{outgoingowl}. 
The network\_stack, as seen in Figure~\ref{fig:network_stack2}, reveals that the private IP given by the smartwatch from outgoingowl was 192.162.35.52. Similar to the leakage using FTP, cross-verification of information obtained from a smartwatch and evidence left on a PC can confirm the connection information with the PC. For example, if a user uses sftp.exe from a PC through a terminal, the investigator can check the private IP and port of the smartwatch in the \texttt{\%UserProfile\%\textbackslash.ssh\textbackslash known\_hosts} file to see that the smartwatch has connected to the PC.

\subsubsection{Act of connecting a smart watch with a hidden camera}
\label{subsubsec:find_hidden_camera}

From the usagestats log in Figure~\ref{fig:usagestats}, it can be seen that the user activated the Lookcam (com.view.ppcs) application around 13:59 on May 9, 2023. According to netstats in Figure~\ref{fig:netstats}, it can be identified that the SSID of the Wi-Fi network that was connected during the active time of the app was F818026FNMEN, and it can be seen that a total of 31MB of data was transmitted. In addition, st=1683547200 was connected at 21:00 on May 8, 2023, and it can be confirmed that data equivalent to about 125MB was transmitted. The reason why usagestats could not check the activation log of the small camera control application around 21:00 on May 8 is that only the last time information used was left in the usagestats log 24 hours after using the Lookcam (com.view.ppcs) application, as explained in Section~\ref{subsubsec:dumpsys_usagestats}.

%% file: _sections/sec_6-Discussion.tex
\section{Discussion}
\label{sec:Discussion}

\subsection{Limitation}
\label{subsec:limitation}

Although Section \ref{subsec:Case_Study} demonstrate that user behavior can be identified through ADB log records on smartwatches, it is important to acknowledge the limitations associated with this method. Firstly, the lifespan of ADB logs poses a significant constraint. In the case of usagestats, if trying to investigate a user's actions 24 hours after a crime, only the last recorded action of the application and the number of uses can be known, making it difficult to determine the precise of value of the action time. In addition, the network\_stack log has a fatal disadvantage which is the volatile characteristic of the log meaning that the log disappears if the smartwatch is rebooted. This issue is magnified by the smartwatch's limited battery capacity, which leaves it susceptible to quick depletion, thus compounding the problem.

Ambiguity presents a significant limitation in the context of Android Debug Bridge (ADB) log analysis, particularly within the domain of netstats logs. These logs record the data size exchanged with connected Wi-Fi networks on an hourly basis. This can also act as a limitation from the perspective of the investigation. When a smartwatch connects to multiple Wi-Fi networks within the same hour, as depicted in Figure~\ref{fig:same_st}, it becomes challenging to make a definite decision which Wi-Fi network corresponds to the private IP identified in the network\_stack log. This ambiguity can complicate investigations and impede accurate attribution of network activities. 

Finally, the traces of connection between PCs and smartphones and data transmission and reception can be clearly demonstrated, but there is a limitation in that it is not possible to know exactly what data has been transmitted and received.

\subsection{Prevention}
\label{subsec:prevention}

To mitigate the risks of technical and management information leakage through smartwatches, it is crucial to implement preventive measures. First, it is necessary to update the use of smartwatches in the company's technology protection regulations. While Large-scale corporations are employing measures such as security stickers and Mobile Device Management (MDM) to safeguard against vulnerabilities that may lead to information leakage through mobile devices. They are also utilizing technical solutions such as company-wide Wi-Fi blocking, and whitelisted network access control mechanisms to protect against the unauthorized dissemination of technological and managerial information. However, compared to the proliferation and potential risk of smartwatches as leakage tools, regulations related to smartwatch security appear to be insufficient. Unauthorized areas and situations for smartwatches should be defined, and if there is an in-house security team, it should be regularly audited.

Although revising and updating technology protection regulations may appear formal, they are critical for maintaining security. But corporate statistics and government statistics reports show that they are not formal and necessary. According to a report by INFOWATCH, leakage by insiders accounts for 63.5\% of the types of technical and management information leakage internationally~\cite{companyreport1}. According to a 2022 survey by the Ministry of SMEs and Startups of Korea, 68.4\% of Korea's technology infringement types were caused by leakage by internal employees, and 21.1\% was caused by third parties (outsourcing, outsourcing services, suppliers, etc.). According to the report, 45\% of small and medium-sized companies in Korea do not have technology protection regulations, and 14.2\% of medium-sized companies~\cite{nationalreport1}. Thus, redefining and updating technology protection regulations, including smartwatches, can raise awareness of leakage among employees in the enterprise, helping to prevent technology and management information leakage crimes, especially in small and medium-sized enterprises.

\begin{figure}[h]
    \begin{center}
    \includegraphics[width=0.48\textwidth]{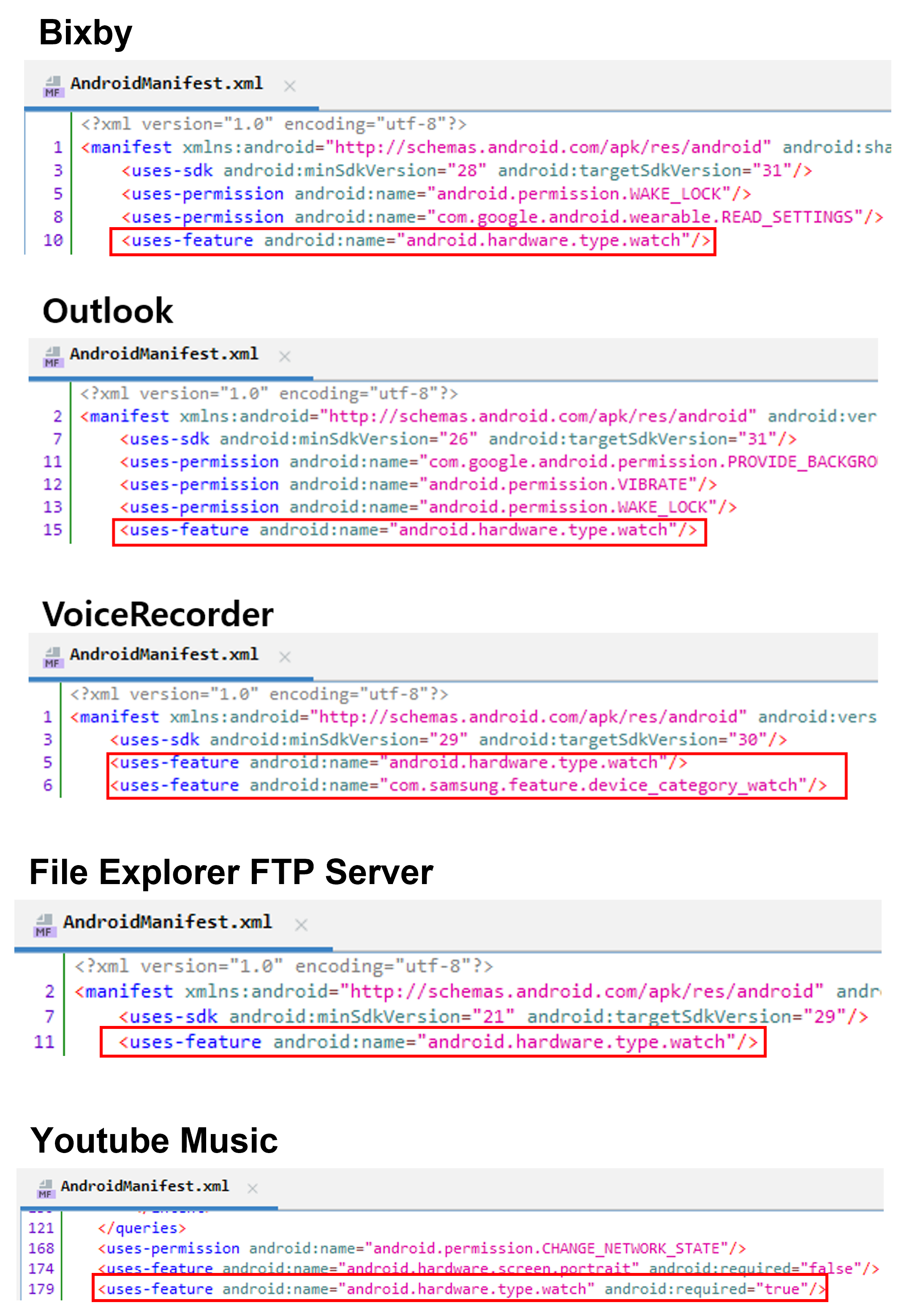}
    \end{center}
    \caption{AndroidManifest.xml of applications for smartwatch}
    \label{fig:androidmanifest}
\end{figure}

For large companies with with robust technical security measure, developing MDM for smartwatches can be suggested as a prevention method. The cases analyzed in the previous case study were a common method of using a Wi-Fi network, and it was confirmed that a method of downloading an application for a smartphone that is not normally installed on a smartwatch was also used. The \texttt{\textless{}uses-feature android:name="android.hardware.type.watch"\textgreater{}} entry in the \texttt{AndroidManifest.xml} file, which can be found in Figure~\ref{fig:androidmanifest}, is information found in applications built for smartwatches. Therefore, it is possible that the absence of the item in the AndroidManifest.xml file of the application installed on the smartwatch is a case where the application for the smartphone is installed on the smartwatch. Leveraging these characteristics, an MDM system for smartwatches can be implemented to restrict WiFi access on company premises, block the installation of applications not designed for smartwatches, and manage permissions, such as microphone access, on smartwatches. These measures can effectively prevent incidents of technical and management information leakage through smartwatches.

%% file: _sections/sec_7-Conclusion.tex

\section{Conclusion}
\label{sec:conclusion}

Smartwatches have become widely accessible to the general public, functioning as small computers on users' wrists. This study was initiated to examine concerns related to the potential exploitation of smartwatches for committing crimes, particularly by leveraging the developer-friendly features of the Android-based Wear OS operating systems. To demonstrate this, experiments were conducted using the Galaxy Watch 4 and 5 series operating with the Wear OS Powered by SAMSUNG operating system. By installing currently distributed smartwatch-exclusive applications and smartphone-exclusive applications onto the smartwatch via ADB commands, information leakage is possible through the use of FTP, SFTP, hidden cameras, and other means.

Subsequently, a log analysis method utilizing the ADB dumpsys command was presented to aid in investigating crimes involving smartwatches. The dumpsys usagestats command allowed for the examination of precise application operation records for a 24-hour period, as well as approximate usage records for a week, month, and year. However after a 24-hour pereiod elapses, it was possible to only confirm the last execution time and the total number of executions. Through the dumpsys netstats command,  it was possible to access information related to the smartwatch's network connections, including the ability to identify the connection times of previously connected Wi-Fi networks on an hourly basis, as well as ascertain the SSID and the size of the transmitted and received data. The dumpsys network\_stack command provided DHCP communication information to and from the wired Wi-Fi network and the private IP number given.

Various information on app usage records and Wi-Fi connections could be examined, but there was a limitation in that the analysis became ambiguous over time due to the short lifespan and period of logs that can be obtained through the dumpsys command. In the case of the existing Android smartphone Forensics, it is possible to connect through a USB port and image using vulnerabilities, so precise analysis can be performed, but the latest OS's smartwatch does not have a USB port, which limits the directory that can be investigated. Currently, Android-based Wear OS smartwatches are as difficult to achieve as accurate forensics as Android smartphones.

Although technology leakage crimes through smartwatches may not be prevalent at present, advancements in smartwatch capabilities and increasing functionalities warrant attention. As technology advances, the performance of smartwatches will be upgraded and functions will be diversified. Currently, smartwatches are often out of the scope of security policies. Many institutions still do not control smartwatches when controlling access, and in the event of a technology leak crime, e-mail, cloud, and mobile storage media are subject to investigation, but there are not many situations where smartwatches are considered. Therefore, attempts should be made to judge the security threat of smartwatches, which are currently undervalued compared to performance, and reflect them in security policies. Furthermore, there is an imperative need for meticulous research into vulnerabilities that permit a Full File System analysis of the smartwatch. This research would facilitate an exhaustive understanding of potential security weaknesses. Concurrently, additional technological advancements are required, particularly in the development of mobile data acquisition tools, to allow for wireless connectivity and debugging functionalities. Such enhancements would ease the process of smartwatch data retrieval and thereby contribute to a more robust and precise investigative methodology.